\documentstyle[preprint,prb,aps,epsfig,amssymb,amsbsy,amsmath,multirow]{revtex}
\newcommand{\etal}{{\it et al.}}
\newcommand{\op}[1]{%
    \fontdimen12\textfont3=2pt\fontdimen12\scriptfont3=1.4pt%
    \!\null\mathop{\vphantom{#1}\smash{#1}}\limits_{\sim}\null\!}
\newcommand{\vek}[1]{{\!\vec{\,#1}}}
\newcommand{\xref}[1]{\protect\ref{#1}}
\newcommand{\figref}[1]{Fig.~\protect\ref{#1}}
\newcommand{\fmref}[1]{(\protect\ref{#1})}
\def\ket#1{\, | \, {#1} \, \rangle}

\begin{document}
\tightenlines
\draft
\title{Rotational modes in molecular magnets with 
antiferromagnetic Heisenberg exchange}
\author{J. Schnack}  
\address{Universit\"at Osnabr\"uck, Fachbereich Physik \\  
         Barbarastr. 7, 49069 Osnabr\"uck, Germany}
\author{Marshall Luban}  
\address{Ames Laboratory \& Department of Physics and Astronomy,
Iowa State University\\ Ames, Iowa 50011, USA}
\maketitle
\begin{abstract}
In an effort to understand the low temperature behavior of
recently synthesized molecular magnets we present numerical
evidence for the existence of a rotational band in systems of
quantum spins interacting with nearest-neighbor
antiferromagnetic Heisenberg exchange. While this result has
previously been noted for ring arrays with an even number of
spin sites, we find that it also applies for rings with an odd
number of sites as well as for all of the polytope
configurations we have investigated (tetrahedron, cube,
octahedron, icosahedron, triangular prism, and axially truncated
icosahedron). It is demonstrated how the rotational band levels
can in many cases be accurately predicted using the
underlying sublattice structure of the spin array.  We
illustrate how the characteristics of the rotational band can
provide valuable estimates for the low temperature magnetic
susceptibility.

\end{abstract}
\widetext
\section{Introduction and summary}

The subject of molecular magnets has greatly advanced in recent
years due to notable progress in synthesizing bulk samples of
identical molecular-size units
\cite{SGC:Nat93,GCR94,Gat:AM94,Cor:NATO96}, each containing a
relatively small number of paramagnetic ions (``spins") that
mutually interact via Heisenberg exchange. An especially
attractive feature of many of these systems is that the
intermolecular magnetic interactions are utterly negligible as
compared to the intramolecular interactions.

Already at this early stage it is clear that even Heisenberg
systems of relatively modest size pose a major theoretical
challenge. A stunning example is provided by the recently
synthesized molecular magnet \cite{MSS:ACIE99} 
\{Mo$_{72}$Fe$_{30}$\}, where the 30 Fe$^{3+}$ ions (spins 5/2)
occupy the sites of an icosidodecahedron.  The total dimension
of the Hilbert space for this spin system is a staggering
$6^{30}$, namely of order Avogadro's number, utterly precluding
the calculation of the energy eigenvalues and eigenvectors on any
imagined configuration of immense, ultra-fast computers. This is
the context for our exploration in this article of a generic
feature of the low-lying excitation energies of a finite number
of spins interacting via antiferromagnetic Heisenberg
exchange. 
With the knowledge of the low-lying excitations one can establish the 
very low temperature properties and in some cases even arrive at an estimate
of the temperature range for the manifestation of essentially
quantum behavior. This is illustrated in the present article
for \{Mo$_{72}$Fe$_{30}$\}; we arrive at an expression for the temperature
dependence of the weak-field susceptibility at very low temperatures.
We also provide an
estimate for the temperature above which this
system can reliably be described by
the classical Heisenberg model, a far more practical theoretical
platform than the corresponding quantum model. Indeed the
classical Heisenberg model is currently being exploited so as to
provide detailed quantitative predictions \cite{Nature,ScL} for
\{Mo$_{72}$Fe$_{30}$\} that are being compared to the results of
ongoing experiments.\cite{Nature}

In this article we focus on a generic feature of highly
symmetric geometric arrays (ring structures or polytopes)
housing a finite number of spins interacting via
antiferromagnetic Heisenberg exchange.  Whatever the intrinsic
spin of the individual paramagnetic ions, or the specific
geometrical symmetries of the spin array, there always exist
what we refer to as rotational modes.  Besides their intrinsic
interest, we show that the knowledge of these modes can in many
cases be used to obtain good estimates of physical observables,
such as the magnetic susceptibility at very low temperatures.
We consider in the following a finite number, $N$, of quantum
spins, each of intrinsic spin $s$, that in most examples
interact via nearest-neighbor
isotropic Heisenberg exchange. We assume that all nearest-neighbor pairs of
spins interact with the same coupling constant, one which favors
antiferromagnetic ordering.  If $\op{\vek{S}}$ denotes the total
spin operator, the operator $\op{\vek{S}}^2$ commutes with the
Hamilton operator, and thus we can structure the set of all
energy eigenvalues according to the total spin quantum number
$S$, extending up to $S_{\text{max}}=N s$ . A key point of the
present work relates to the subset of minimal energies for the
allowed values of $S$.  We may summarize our findings as
follows: {\it Whatever the details of the system, this subset of
minimal energies appears to define what we shall refer to as a
``rotational band", i.e., is well approximated by a dependence
on $S$ of the form $S (S+1)$.}  We choose the term ``rotational
band" to indicate that this portion of the spectrum is similar
to that of a rigid rotor. Similar behavior is commonly found also in
nuclear and atomic physics.

The occurrence of a rotational band has been noted on several
occasions for an even number of spins defining a ring structure.
The minimal energies have been described
\cite{TDP:JACS94,LGC:PRB97A,LGC:PRB97B,ACC:ICA00} as ``following
the Land\'{e} interval rule".  However, we find that the same
property also occurs for rings with an odd number of spins as
well as for the various polytope configurations we have
investigated, in particular for quantum spins positioned on the
vertices of a tetrahedron, cube, octahedron, icosahedron,
triangular prism, and an axially truncated icosahedron.
Rotational modes have also been found in the context of finite
triangular lattices of spin-$1/2$ Heisenberg antiferromagnets
\cite{BLL:PRB94,GSS:PRB89}.

Using only the sublattice structure of the various spin arrays,
which is provided by symmetry arguments, we are able to
approximate the coefficient of the $S (S+1)$ dependence to good
accuracy. Our method describes in general how this approximate
coefficient can be deduced. We can therefore obtain an estimate
of the ground state energy as well as the low-lying rotational
excitations.  It is clear that at low temperatures these minimal
energies provide the major contribution to thermal averages.
This enables us to discuss the low-temperature behavior of
quantities such as the magnetic susceptibility without knowledge
of the complete eigenvalue spectrum.  We illustrate these
considerations for the special case of \{Mo$_{72}$Fe$_{30}$\}.

The layout of this article is as follows. In Sec. \xref{sec-2}
we present our numerical findings for various Heisenberg spin
systems and motivate in Sec. \xref{sec-3} how the rotational
band is connected to the topology of spin sites. Finally in
Sec. \xref{sec-4} we discuss some implications of the rotational
band on physical observables.

\section{Rotational bands}
\label{sec-2}

The Hamilton operator for the isotropic Heisenberg model 
in the absence of an external magnetic field reads 
\begin{eqnarray}
\label{E-1-1}
\op{H}
&=&
-
2\,J\,
\sum_{(u,v)}\;
\op{\vek{s}}(u) \cdot \op{\vek{s}}(v)
\ ,\quad \forall u: s(u)=s
\ ,
\end{eqnarray}
where $J$ is the exchange interaction with units of energy, and
$J<0$ results in antiferromagnetic coupling. The vector
operators $\op{\vek{s}}(u)$, underlined with a tilde, are the
single-particle spin operators with eigenvalue equations
\begin{eqnarray}
\label{E-1-2}
\left(\op{\vek{s}}(u)\right)^2\, \ket{s(u)\; m(u)}
&=&
s(s+1)\, \ket{s(u)\; m(u)}
\\
\op{{s}}_z(u)\, \ket{s(u)\; m(u)}
&=&
m(u)\, \ket{s(u)\; m(u)}
\nonumber
\ .
\end{eqnarray}
The sum in \eqref{E-1-1} runs over all distinct interacting
pairs $(u,v)$ of spins at positions $u$ and $v$. For a closed
ring with nearest-neighbor interaction the index $v$ would
simply equal $u+1$ and the sum is understood to fulfill the
cyclic boundary condition.

\subsection{Heisenberg square}

\begin{figure}[t]
\begin{center}
\epsfig{file=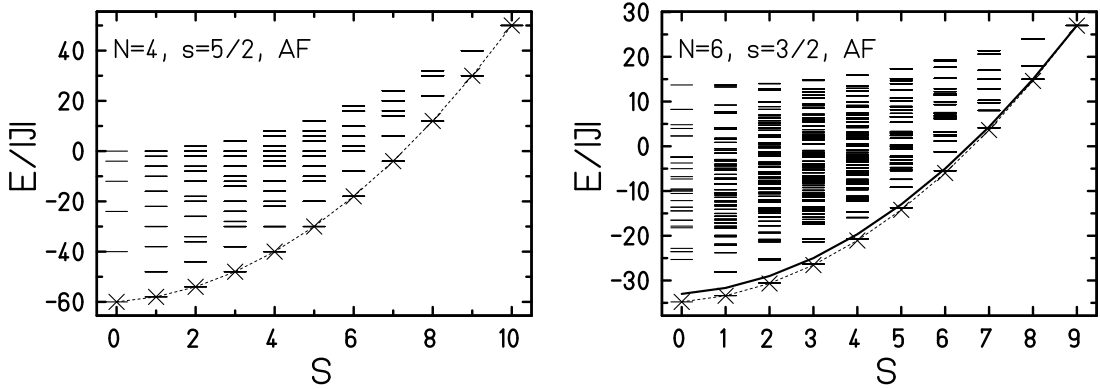,width=145mm}
\vspace*{1mm}
\caption[]{Energy spectra of antiferromagnetically coupled
Heisenberg spin rings (horizontal dashes).
Here and in all subsequent figures the crosses 
connected by the dashed line always 
represent the fit to the rotational band according to
\eqref{E-2-13}, which by definition matches both the lowest 
and the highest energies
exactly. 
On the l.h.s the dashed line reproduces the exact rotational band, 
whereas on the
r.h.s. it only approximates it,
but to high accuracy. The solid line on the r.h.s. corresponds to
the approximation of \eqref{E-2-7}.}
\label{F-2-1}
\end{center} 
\end{figure}

One of the few systems that possesses a rigorous parabolic
rotational band is the Heisenberg square, i.e., a ring with
$N=4$. Because the Hamilton operator can be rewritten as
\begin{eqnarray}
\label{E-2-1}
\op{H}
=
-
J\,
\left(
\op{\vek{S}}^2 - \op{\vek{S}}_{13}^2 -\op{\vek{S}}_{24}^2
\right)
\ , \
\op{\vek{S}}_{13}=\op{\vek{s}}(1)+\op{\vek{s}}(3)
\ , \
\op{\vek{S}}_{24}=\op{\vek{s}}(2)+\op{\vek{s}}(4)
\ ,
\end{eqnarray}
with all spin operators $\op{\vek{S}}^2$, $\op{\vek{S}}_{13}^2$
and $\op{\vek{S}}_{24}^2$ commuting with each other and with
$\op{H}$, one can directly obtain the complete set of
eigenenergies and these are characterized by the quantum numbers
$S$, $S_{13}$ and $S_{24}$. In particular, the lowest energy for
a given total spin quantum number $S$ occurs for the choice
$S_{13}=S_{24}=2s$
\begin{eqnarray}
\label{E-2-2}
E_{S, min}
=
-
J\,
\left[
S\,(S+1) - 2\cdot 2 s\, (2s+1)
\right]
=
E_0 - J\,S\,(S+1)
\ ,
\end{eqnarray}
where $E_0=4 s (2s+1) J$ is the exact ground state energy.  The
various energies $E_{S, min}$ form a rigorous parabolic rotational band
of excitation energies.  Therefore, these energies coincide with
a parabolic fit (crosses connected by the dashed line on the
l.h.s. of \figref{F-2-1}) passing through the antiferromagnetic
ground state energy and the highest energy level, i.e., the
ground state energy of the corresponding ferromagnetically
coupled system.

\subsection{Heisenberg rings with $N>4$}

\begin{figure}[t]
\begin{center}
\epsfig{file=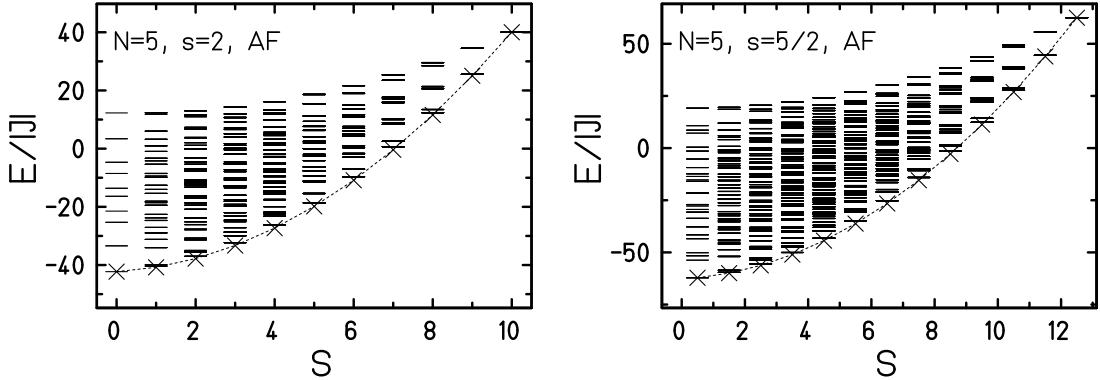,width=145mm}
\vspace*{1mm}
\caption[]{Energy spectra of antiferromagnetically coupled
Heisenberg spin rings with $N=5$, $s=2$ (l.h.s.) and $s=5/2$ (r.h.s.).}
\label{F-2-2}
\end{center} 
\end{figure}

We have calculated all energy levels by diagonalizing
\cite{BSS:JMMM00} the Hamilton 
matrix for a variety of rings with different values of $N$
and $s$.  All of these systems exhibit a rotational band,
irrespective whether $N$ is even or odd and for both integer and
half-integer values of $s$.
That is, the subset of minimal energies is well approximated
by a dependence on $S$ which is proportional to $S(S+1)$,
i.e., follows the Land\'{e} interval rule
\cite{TDP:JACS94,LGC:PRB97A,LGC:PRB97B,ACC:ICA00},
\begin{eqnarray}
\label{E-2-13}
E_{S, min}
\approx
E_a
- J\, \frac{D(N,s)}{N}\,
S (S+1)
\ .
\end{eqnarray}
We determine the parameters $E_a$ and $D(N,s)$ so that formula
\fmref{E-2-13} reproduces our calculated values of the lowest
and highest energies of the rotational band, i.e., the ground
state energy of the antiferromagnetic system and the ground
state energy of the corresponding ferromagnetic system.  In all
cases we have observed, that if deviations occur the fitting
parabola of \fmref{E-2-13} lies below the rotational band.  Use
of \fmref{E-2-13} is illustrated in \figref{F-2-1} (r.h.s.) for
the case $N=6$ and $s=3/2$. The figure shows the complete
spectrum (horizontal dashes) as well as the fit according to
\eqref{E-2-13} (crosses connected by a dashed line).  One
observes that the fit very nearly matches the energies of the
rotational band, meaning that the Land\'{e} interval rule is
obeyed with high accuracy. As a second example, spectra are
shown for rings of five spins with $s=2$ (\figref{F-2-2},
l.h.s.) and $s=5/2$ (\figref{F-2-2}, r.h.s.).  Inspecting the
low-lying excitations one notices that the rotational band for
odd rings is not separated from the remaining states as much as
it is for even rings. This remark also pertains to other, larger
odd values of $N$.

In Table \xref{T-1} we collect the coefficients $D(N,s)$ for the
rings we have investigated. In all cases $D(N,s)\approx 4$.
For the odd rings the values of $D(N,s$ 
may be somewhat smaller than four.  We
will dwell on this fact in Sec. \xref{sec-3}.

It should be noted that in the large-$N$-limit the rotational
levels \fmref{E-2-13} become degenerate since $D(N,s)$ remains
finite. Therefore, excitations within the rotational band should
not be confused with magnons.

\subsection{Frustrated spin rings}

\begin{figure}[t]
\begin{center}
\epsfig{file=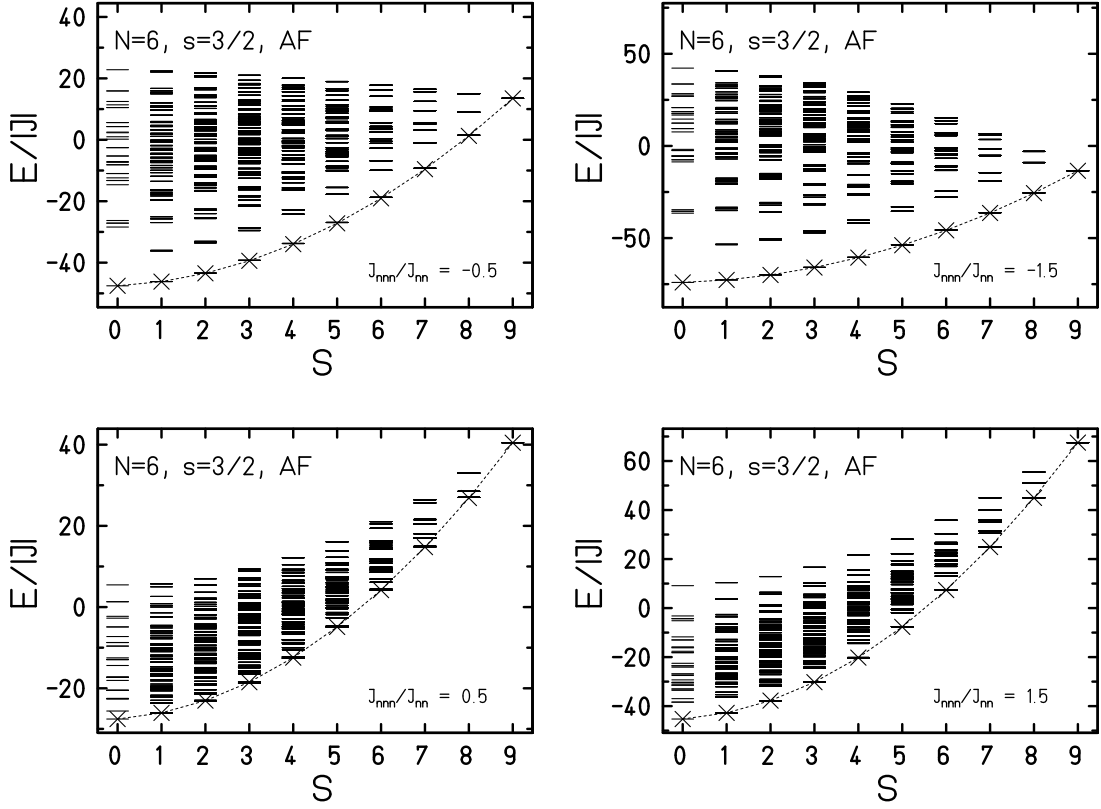,width=145mm}
\vspace*{1mm}
\caption[]{Energy spectrum of an antiferromagnetically next
neighbor coupled Heisenberg ring with competing next-nearest
neighbor interaction.}
\label{F-2-5}
\end{center} 
\end{figure}

Even for spin rings with next-nearest neighbor interaction the
rotational band persists. The energy spectra of \figref{F-2-5}
have been calculated for the Hamilton operator
\begin{eqnarray}
\label{E-2-10A}
\op{H}
&=&
-
2\,J_{nn}\,
\sum_{u=1}^N\;
\op{\vek{s}}(u) \cdot \op{\vek{s}}(u+1)
-
2\,J_{nnn}\,
\sum_{u=1}^N\;
\op{\vek{s}}(u) \cdot \op{\vek{s}}(u+2)
\ ,
\end{eqnarray}
where all spins have been taken to be $s(u)=3/2$.
In \figref{F-2-5} we display spectra for a ring with $N=6$ and
$s=3/2$ for various ratios of the two coupling constants
$J_{nn}$ and $J_{nnn}$.  Although details of the spectra differ,
the overall appearance persists, and in particular the minimal
energies define a rotational band.

\subsection{Heisenberg polytopes}

\begin{figure}[t]
\begin{center}
\epsfig{file=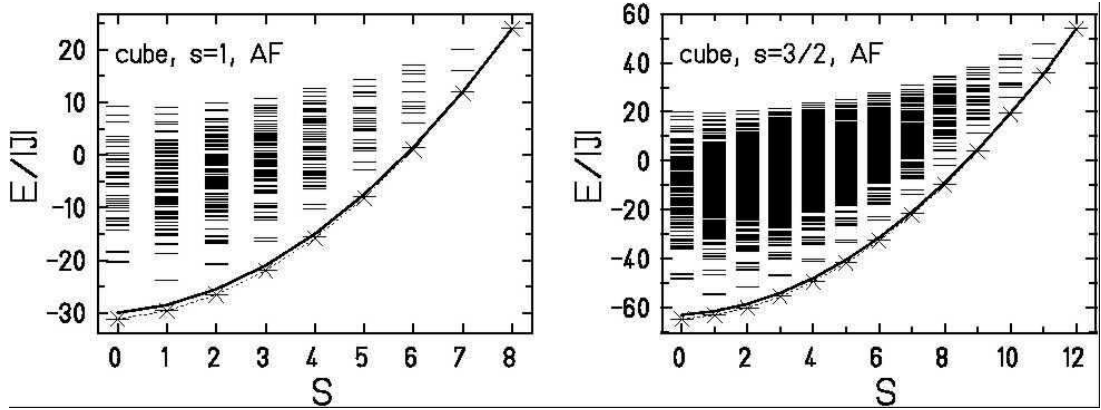,width=145mm}
\vspace*{1mm}
\caption[]{Energy spectrum of antiferromagnetically coupled
Heisenberg cubes. The solid lines correspond to
the approximation of \eqref{E-2-11} and \eqref{E-2-12},
i.e. $D=6$.}
\label{F-2-3}
\end{center} 
\end{figure}

\begin{figure}[t]
\begin{center}
\epsfig{file=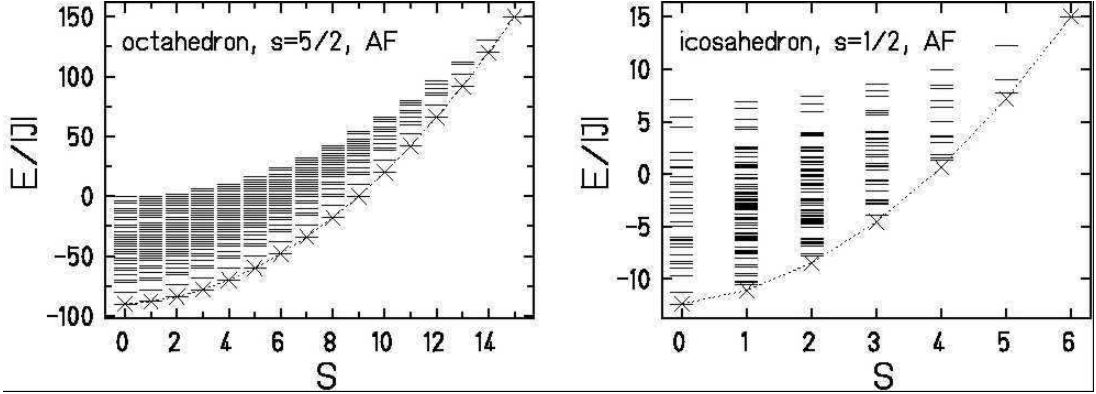,width=145mm}
\vspace*{1mm}
\caption[]{Energy spectrum of an antiferromagnetically coupled
Heisenberg octahedron (l.h.s.) and icosahedron (r.h.s.).}
\label{F-2-4}
\end{center} 
\end{figure}

In order to illustrate the generality of the rotational band we
provide several additional examples, the tetrahedron, the cube
(\figref{F-2-3}), the octahedron (\figref{F-2-4}, l.h.s.), and
the icosahedron (\figref{F-2-4}, r.h.s.). As in the previous
cases the displayed energy eigenvalues are calculated by
numerical diagonalization, except for the tetrahedron and
octahedron which can be solved analytically.

The tetrahedron of spins is a worthy textbook problem; it can be
solved with a few lines of algebra because the Hamilton operator
simplifies to
\begin{eqnarray}
\label{E-2-9B}
\op{H}
=
-
J\,
\left(
\op{\vek{S}}^2 - 4\, \op{\vek{s}}^2
\right)
\ .
\end{eqnarray}
Therefore, the spectrum of this system 
consists exclusively of a rotational band.

The case of the octahedron is similar to the Heisenberg square;
the Hamilton operator can be written as
\begin{eqnarray}
\label{E-2-9}
\op{H}
=
-
J\,
\left(
\op{\vek{S}}^2 - \op{\vek{S}}_{A}^2 - \op{\vek{S}}_{B}^2 - \op{\vek{S}}_{C}^2
\right)
\ ,
\end{eqnarray}
where $\op{\vek{S}}_{A}$, $\op{\vek{S}}_{B}$, $\op{\vek{S}}_{C}$
are the sums for pairs of spins situated at opposite vertices of
the octahedron, and $\op{\vek{S}}$ is the total spin. The spin
operators $\op{\vek{S}}^2$, $\op{\vek{S}}_{A}^2$,
$\op{\vek{S}}_{B}^2$ and $\op{\vek{S}}_{C}^2$ commute with each
other and with $\op{H}$. Thus the eigenvalues of $\op{H}$ may be
written down at once and they are given in terms of the quantum
numbers $S$, $S_{A}$, $S_{B}$ and $S_{C}$. Therefore, the lowest
energy for a given value of $S$ is achieved if
$S_{A}=S_{B}=S_{C}=2s$, and its value is given by
\begin{eqnarray}
\label{E-2-9A}
E_{S, min}
=
-
J\,
\left[
S\,(S+1) - 3\cdot 2 s\, (2s+1)
\right]
\ .
\end{eqnarray}
This is another one of the few cases where 
the minimal energies define a rigorous
rotational band (see \figref{F-2-4}, l.h.s.).

The remaining examples of the cube (\figref{F-2-3}) and the
icosahedron (\figref{F-2-4}, r.h.s.) illustrate the behavior of
the gaps between the rotational band and the remaining
eigenenergies. It is worth noting that the rotational band of
the icosahedron is not as well separated from higher energy
levels as it is for the cube.  This behaviour is similar to that
discussed above for even and odd rings.  Systems which are
bipartite, i.e., can be subdivided into two sublattices with
interactions only between spins of different sublattices (rings
with even $N$ and the cube), show a significant gap, whereas
systems, that are non-bipartite appear to show much smaller
gaps.\cite{VRT:JMMM98} The latter systems are often also called
frustrated.\cite{RIV:JLTP95} Two other cases we have studied,
the equilateral triangle prism and an axially truncated
icosahedron, conform with these trends.

\section{Conjecture on rotational bands}
\label{sec-3}

\subsection{Heisenberg rings; even $N$}

It turns out that an accurate approximate formula for the
coefficient $D(N,s)$ of \eqref{E-2-13} can be developed using
the sublattice structure of the spin array. As an introductory
example we repeat the basic ideas for Heisenberg rings with an
even number of spin sites \cite{ACC:ICA00}.  Such rings are
bipartite and can be decomposed into two sublattices, labeled
$A$ and $B$, with every second spin belonging to the same
sublattice.  From classical spin dynamics it is known that the
classical ground state, sometimes called the classical N\'eel
state \cite{BLL:PRB94}, is given by an alternating sequence of
opposite spin directions. On each sublattice the spins are
mutually parallel.  Therefore, a quantum trial state, where the
individual spins on each sublattice are coupled to their maximum
values, $S_A=S_B=N s / 2$, could be expected to provide a
reasonable approximation to the true ground state, especially if
$s$ assumes large values. Such trial states are called
N\'eel-like. For rings with even $N$ the approximation to the
respective minimal energies for each value of the total spin
$\op{\vek{S}}=\op{\vek{S}}_A + \op{\vek{S}}_B$ is found to be
given by \cite{ACC:ICA00}
\begin{eqnarray}
\label{E-2-7}
E_{S, min}^{\mbox{\scriptsize approx}}
=
- \frac{4\, J}{N}\,
\left[
S (S+1) - 2 \frac{N s}{2} \left( \frac{N s}{2} + 1 \right)
\right]
\ .
\end{eqnarray}
This approximation exactly reproduces the energy of the highest
energy eigenvalue, i.e., the ground state energy of the
corresponding ferromagnetically coupled system ($S=N s$), since
the true eigenstate with all spins assuming their largest $m$
quantum number, $m=s$, is a linear
combination of N\'eel-like states. For all smaller $S$ the
approximate minimal energy $E_{S, min}^{\mbox{\scriptsize
approx}}$ is bounded from below by the true one (Rayleigh-Ritz
variational principle).  The solid curve displays this behavior
for the example of $N=6$, $s=3/2$ in \figref{F-2-1}
(r.h.s.). The entries in Table \xref{T-1} provide additional
numerical support for the approximation $D(N,s)\approx 4$
adopted in \fmref{E-2-7}. For each fixed even $N$ the
coefficient $D(N,s)$ approaches 4 with increasing $s$.

The approximate spectrum, \eqref{E-2-7}, is similar to that 
of two spins, 
$\op{\vek{S}}_A$ and $\op{\vek{S}}_B$,
each of spin quantum number $N s / 2$, that
are coupled by an effective interaction of 
strength $4 J/N$. Therefore, one can
equally well say, that the approximate rotational band
considered in \fmref{E-2-7} is associated with 
an effective Hamilton operator 
\begin{eqnarray}
\label{E-2-6}
\op{H}^{\mbox{\scriptsize approx}}
&=&
- \frac{4\, J}{N}\,
\left[
\op{\vek{S}}^2 - \op{\vek{S}}_A^2 - \op{\vek{S}}_B^2
\right]
\ ,
\end{eqnarray}
where the two sublattice spins, $\op{\vek{S}}_A,
\op{\vek{S}}_B$, assume their maximal value $S_A=S_B=N s /
2$. This result agrees with that obtained by a different
procedure, which utilizes a Fourier representation of the
Hamilton operator.\cite{BLL:PRB94}

In retrospect one realizes that all we needed for rings with
even $N$ to arrive at \fmref{E-2-7} was a sublattice structure
in order to build N\'eel-like trial states.  The sublattice
structure can be deduced from the classical ground state or the
symmetries of the spin array, which manifest themself in the
classical ground state. In the case of rings with an even number
of sites the symmetry is the cyclic shift symmetry: Sublattice
$A$ can be transformed into sublattice $B$ by a single shift.
These considerations lead us in the following subsection to a
generalization for other systems.

\subsection{Conjecture}

We assume that the Heisenberg spin system can be decomposed into
$N_{sl}$ sublattices according to a symmetry transformation.
The spins of each sublattice are to be coupled to their maximal
values $S_{sl}=N s/N_{sl}$.  Then we conjecture that the
energies of the rotational band can be approximated as
\begin{eqnarray}
\label{E-2-11}
E_{S, min}^{\mbox{\scriptsize approx}}
=
- J\, \frac{D}{N}\,
\left[
S (S+1) - 
N_{sl}\;
S_{sl} (S_{sl}+1)
\right]
\ ,
\end{eqnarray}
where the parameter $D$ is to be fixed by the requirement that
the energy $E_{\mbox{\scriptsize ferro}}=- 2 J\, N_b\, s^2$ of
the corresponding ferromagnetic ground state, for which $S=N s$,
is reproduced exactly.  The energy $E_{\mbox{\scriptsize
ferro}}$ depends only on the number of distinct bonds $N_b$.
Therefore, the coefficient $D$ is given by
\begin{eqnarray}
\label{E-2-12}
D
&=&
2\, \frac{N_b}{N}\,
\frac{1}{1 - \frac{1}{N_{sl}}}
\ ,
\end{eqnarray}
which is independent of $s$.  For rings with an even number of
sites this formula reproduces the value $D=4$, whereas for the
cube, which has $N_b=12$, it gives $D = 6$. That these values of
$D$ provide a very good estimate of the coefficients $D(N,s)$ is
demonstrated in \figref{F-2-1} (r.h.s.)  for a ring with $N=6$
and $s=3/2$, and in \figref{F-2-3} for cubes with $s=1$ and
$s=3/2$.

If there are several possible partitions into sublattices
according to various symmetries of the spin array one can safely
choose that one which leads to the lowest approximate ground
state energy, because this energy is bounded from below by the
true ground state energy.  For spin arrays with low symmetry the
gained approximation can be rather poor. This is for instance
the case for rings of odd $N$.  The classical ground state
structure would suggest as many sublattices as spin sites, which
leads to rather small coefficients $D$ as depicted in Table
\xref{T-1} (``simple conjecture").  A closer inspection shows
that it is possible to obtain a better approximation if one
defines the approximate rotational band by the parabola passing
through the classical N\'eel ground state energy level and
through the ground state energy level of the corresponding
ferromagnetic system. For rings of odd $N$ the classical N\'eel
ground state energy is
\begin{eqnarray}
\label{E-2-31}
E_{\text{N\'eel}}
&=&
-2\, J\, N\, s^2 \cos\left((N-1)\pi/N  \right)
,
\end{eqnarray}
as can be shown using spin coherent states.\cite{Lie:CMP73}
This leads to the ``refined conjecture"
for odd rings, see Table \xref{T-1}.
For other spin arrays, such as
the icosahedron, it may be simply impossible to derive a good
approximation for the coefficient $D(N,s)$, nevertheless, the
minimal energies always form a rotational band according to
\eqref{E-2-13}.

Using \fmref{E-2-11} and \fmref{E-2-12} one can find an
approximation for the rotational band for larger polytopes such
as the icosidodecahedron, which characterizes the sites of the
30 paramagnetic Fe$^{3+}$ ions in the recently synthesized
molecular magnet \cite{MSS:ACIE99} \{Mo$_{72}$Fe$_{30}$\}.  The
related Hilbert space has a dimension of $(2 s + 1)^N=6^{30}$
which is of the order of Avogadro's number.  Numerical
diagonalization of the Hamilton matrix is totally out of reach.
Nevertheless, we can estimate both the ground state energy and
the form of the rotational band.  The icosidodecahedron,
consisting of 20 triangles and 12 pentagons, has threefold
rotational symmetry and thus three sublattices.  Assuming
nearest neighbor interaction, we have $N_b=60$. Thus, one ends
up with
\begin{eqnarray}
\label{E-2-21}
E_{S, min}^{\mbox{\scriptsize approx}}
=
-\frac{J}{5}\,
S (S+1) 
+
60 J s \left(s+\frac{1}{10}\right)
\ .
\end{eqnarray}
The threefold rotational symmetry of the spin array is also
reflected by the structure of the classical ground state, thus
it is no surprise that quantum and classical N\'eel ground state
energies,
\begin{eqnarray}
\label{E-2-22}
E_{0, min}^{\mbox{\scriptsize approx}}
=
60 J s \left(s+1/10\right)
\quad , \qquad
E_{\text{N\'eel}}
=
60 J s^2
\ ,
\end{eqnarray}
differ from each other only by 4\% for the rather high $s=5/2$.

\section{Low-temperature thermodynamic properties}
\label{sec-4}

It is obvious that at low temperatures the rotational band
energies provide the dominant contribution to thermal averages,
especially if these energies are well separated from the
remaining energy levels. This suggests, for example, that we
approximate the partition function by
\begin{eqnarray}
\label{E-3-3R}
Z(\beta)
&\approx&
\sum_{S=S_{min}}^{S_{max}} 
\, d_S\, e^{-\beta 
\left[ E_a
- J\, \frac{D}{N}\,S (S+1)
\right]}
\ ,
\end{eqnarray}
where $d_S$ is the degeneracy factor of the eigenenergies
belonging to the rotational band, and $E_a=J (D/N)\;N_{sl}
S_{sl} (S_{sl}+1)$ according to \eqref{E-2-11}.  For bipartite
systems, i.e. systems that can be subdivided into two
sublattices according to the theorem of the Lieb-Schultz-Mattis
\cite{LSM:AP61,LiM:JMP62}, $d_S = 2 S + 1$.  For non-bipartite
systems not much is known about the degeneracy.  First
investigations show that it is possible to establish rules for
the degeneracy of certain states also for non-bipartite
systems.\cite{Schnack:PRB00} For the approximate rotational
energies we take the degeneracies to be that resulting from the
coupling of the sublattice spins.  \cite{BLL:PRB94} It might be
that the full Hamiltonian lifts this degeneracy somewhat,
nevertheless this procedure provides a reasonable approximation
for the degeneracies of the true minimal energies.

\begin{figure}[t]
\begin{center}
\epsfig{file=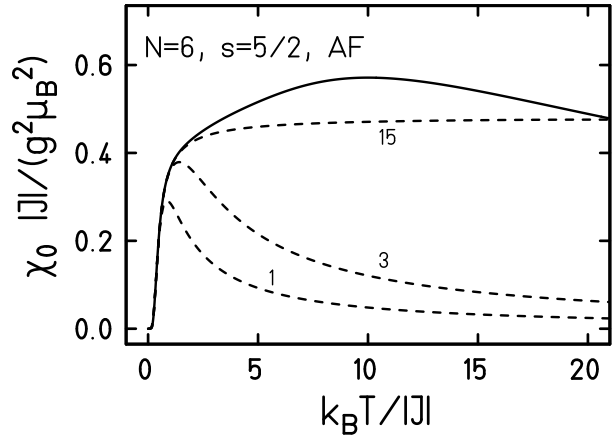,width=80mm}
\vspace*{1mm}
\caption[]{
Zero-field magnetic susceptibility of the
antiferromagnetically coupled Heisenberg spin ring with $N=6$
and $s=5/2$. The solid curve displays the full quantum solution;
the dashed curves correspond to approximation
\eqref{E-3-31}
upon including the contributions of rotational levels up to
$S=1$, $S=3$, and $S=15$.}
\label{F-3-6}
\end{center} 
\end{figure}

The corresponding approximation of the zero-field susceptibility
is then given by
\begin{eqnarray}
\label{E-3-31}
\chi_0
&\approx&
\frac{g^2 \mu_B^2 \beta}{Z}\,
\sum_{S=S_{min}}^{S_{max}} 
\frac{d_S}{2 S + 1}
\left(
\sum_{M=-S}^{M=S} M^2
\right)\,
e^{-\beta 
\left[ E_a
- J\, \frac{D(N,s)}{N}\,S (S+1)
\right]}
\ .
\end{eqnarray}
Inspecting \figref{F-3-6} one sees how the rotational band
contributes to the susceptibility in the case of the ring with
$N=6$ and $s=5/2$.  The rise at low temperatures is mostly
determined by the first excited level of the rotational band
(dashed curve labeled 1).

\begin{figure}[t]
\begin{center}
\epsfig{file=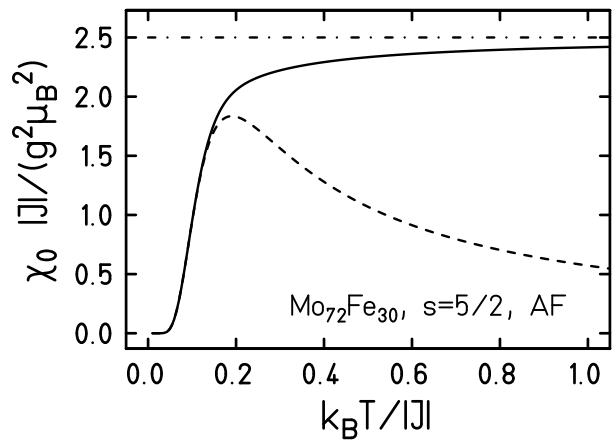,width=80mm}
\vspace*{1mm}
\caption[]{Zero-field magnetic susceptibility of 
\{Mo$_{72}$Fe$_{30}$\} calculated using
\fmref{E-2-21} and \fmref{E-3-31}. The dashed curve is obtained
using only the two lowest levels and the solid curve
using all levels of the approximate 
rotational  band. The classical result is given by the
dash-dot line.}
\label{F-10}
\end{center} 
\end{figure}

As our final example we show in \figref{F-10} the result using
\fmref{E-2-21} and \fmref{E-3-31} for the low temperature
behavior of the zero-field susceptibility for
\{Mo$_{72}$Fe$_{30}$\}.  
For this system $d_s=\text{min}\{(2 S+1)^2,(2 S+1)(76-S)\}$.
The susceptibility (solid curve) rises
very rapidly with increasing temperature to the result
\cite{AxL} for the classical Heisenberg model (dash-dot
curve). Inspecting \eqref{E-2-21} one can understand that the
rapid rise is due to the small energy difference between ground
and first excited state and in particular because of the small
coefficient $J/5$ of the $S(S+1)$ term.  In addition the $S=1$
level of the approximate rotational band is 9-fold degenerate
and thus the rapid rise in $\chi_0$ commences at very low
temperatures. Thus, for \{Mo$_{72}$Fe$_{30}$\}, which has a
nearest neighbor coupling constant \cite{Nature} of
$J/k_B\approx 0.75$~K, we expect, that weak-field susceptibility
measurements will confirm the rapid decrease on cooling, which
is a genuine quantum feature and not present in the classical
counterpart, only at temperatures below $T\approx 0.15$~K.


\section*{Acknowledgments}

The authors thank A.~Cornia (Modena), D.~Gatteschi (Florence),
and M.~Karbach (Wuppertal) for helpful discussions and
J.~Richter (Magdeburg) for drawing their attention to the work
of C.~Lhuillier.  We are deeply indebted to our collaborator
C.~Schr\"oder (Telelogic GmbH, Bielefeld) for innumerable
discussions and quantitative results. We also thank DAAD and
NSF for supporting a mutual exchange program.  The Ames
Laboratory is operated for the United States Department of
Energy by Iowa State University under Contract
No. W-7405-Eng-82.



\begin{table}[t]
\begin{center}
\begin{tabular}{|c||c|c|c|c|c|c|l|}
\hline
$s$&\multicolumn{6}{c|}{$N$}&\\
     & 5 & 6 & 7 & 8 & 9 & 10 &\\
\hline
\hline
    & 2.5  & 4 & 2.333  & 4 & 3 & 4 & simple\\
    &      &   &        &   &   &   & conjecture\\
\hline
    & 3.618  & 4 & 3.802 & 4 & 3.879 & 4 & refined\\
    &      &   &        &   &   &   & conjecture\\
\hline
\hline
$\frac{1}{2}$ & 3.8975 & 4.3028 & 4.2982 & 4.5209 & 4.5355 & 4.6770 &\\
\hline
$1$ & 3.8437  & 4.1764 & 4.1430 & 4.2971 & 4.2957 & 4.3809 & \\
\hline
$\frac{3}{2}$ & 3.7591 & 4.1190 & 4.0174 & 4.1977 & 4.1443 & 4.2514 & \\
\hline
$2$ & 3.7399 & 4.0896 & 3.9772 & 4.1482 & 4.0886 & 4.1881 & \\
\hline
$\frac{5}{2}$ & 3.7103 & 4.0718 & 3.9368 & 4.1185 & 4.0420 & 4.1500 & \\
\hline
\end{tabular}
\vspace*{5mm}
\end{center}
\caption[]{
Coefficients $D(N,s)$ for various Heisenberg rings, to be used
in conjunction with \eqref{E-2-13}. 
}\label{T-1}
\end{table}

\end{document}